\documentclass{eas}
\usepackage{graphicx}
\begin{document}

\title{Local Magnetic Field Role in Star Formation} 
\author{Patrick M. Koch}\address{Academia Sinica Institute of Astronomy and Astrophysics, Taipei, Taiwan}
\author{Ya-Wen Tang}\sameaddress{1}
\author{Paul T.P. Ho}\sameaddress{1}
            \secondaddress{East Asian Observatory, 660 N. Aohoku Place, University Park, Hilo, Hawaii 96720, USA}
\author{Qizhou Zhang}\address{Harvard-Smithsonian Center for Astrophysics, Cambridge, MA 02138, USA}
\author{Josep M. Girart}\address{Institut de Ci\`{e}ncies de l'Espai, 
Facultat de Ci\`{e}ncies, C5p 2, 08193 Bellaterra, Catalonia}
\author{Huei-Ru V. Chen}\address{Inst. of Astronomy and Dep. of Physics, National Tsing Hua University, 
Hsinchu, Taiwan}\sameaddress{1}
\author{Shih-Ping Lai}\sameaddress{5,1}
\author{Hua-Bai Li}\address{Department of Physics, The Chinese University of HongKong}
\author{Zhi-Yun Li}\address{Department of Astronomy, University of Virginia, Charlottesville, VA 22904, USA}
\author{Hau-Yu B. Liu}\sameaddress{1}
\author{Marco Padovani}\address{Laboratoire Univers et Particules de Montpellier, Universit\'e de Montpellier, France}
                                    \secondaddress{INAF-Osservatorio Astrofisico di Arcetri, Firenze, Italy}
\author{Keping Qiu}\address{School of Astronomy and Space Science, Nanjing University, Nanjiing 210093, China}
\author{Ramprasad Rao}\address{Academia Sinica Institute of Astronomy and Astrophysics, Hilo, HI 96720, USA}
\author{Hsi-Wei Yen}\sameaddress{1}
\author{Pau Frau}\address{Observatorio Astron\'{o}mico Nacional, E-28014 Madrid, Spain}
\author{How-Huan Chen}\sameaddress{3}
\author{Tao-Chung Ching}\sameaddress{3,5}

\runningtitle{Koch \etal:  Local Magnetic Field Role}
\begin{abstract}
We highlight distinct and systematic observational features of magnetic field morphologies in 
polarized submm dust continuum. We illustrate this with specific examples and show statistical trends from
a sample of 50 star-forming regions.
\end{abstract}
\maketitle
\section{Introduction:  Angle $\delta$}

Dust polarization observations in the mm/submm continuum reveal plane-of-sky
projected magnetic field morphologies.  Here, dust grains are expected to be aligned 
with their shorter axes parallel to the magnetic field, 
thus the detected polarized emission is perpendicular to the B field. 
Despite the recent growing
observations, e.g., CARMA (Hull {\em et al.\/} \cite{hull14}), 
SMA (Zhang {\em et al.\/} \cite{zhang14}), magnetic field strengths can unfortunately 
not be directly inferred from these observed magnetic field morphologies.  Through 
a series of publications (Koch {\em et al.\/} \cite{koch12a}, \cite{koch12b}, \cite{koch13}, \cite{koch14})
we introduced $\delta$ -- the local angle between
an observed magnetic field orientation and the gradient of the underlying
dust continuum Stokes $I$ emission (Figure \ref{example_delta}). With the help of a 
magnetohydrodynamics (MHD) force equation, $\delta$ can be interpreted as a magnetic field
alignment factor, where $\sin|\delta|$ 
measures the fraction of the magnetic field tension force that is directed along an emission (density)
gradient. 

\section{Results}

\begin{figure}
\begin{center}
\includegraphics[scale=0.25]{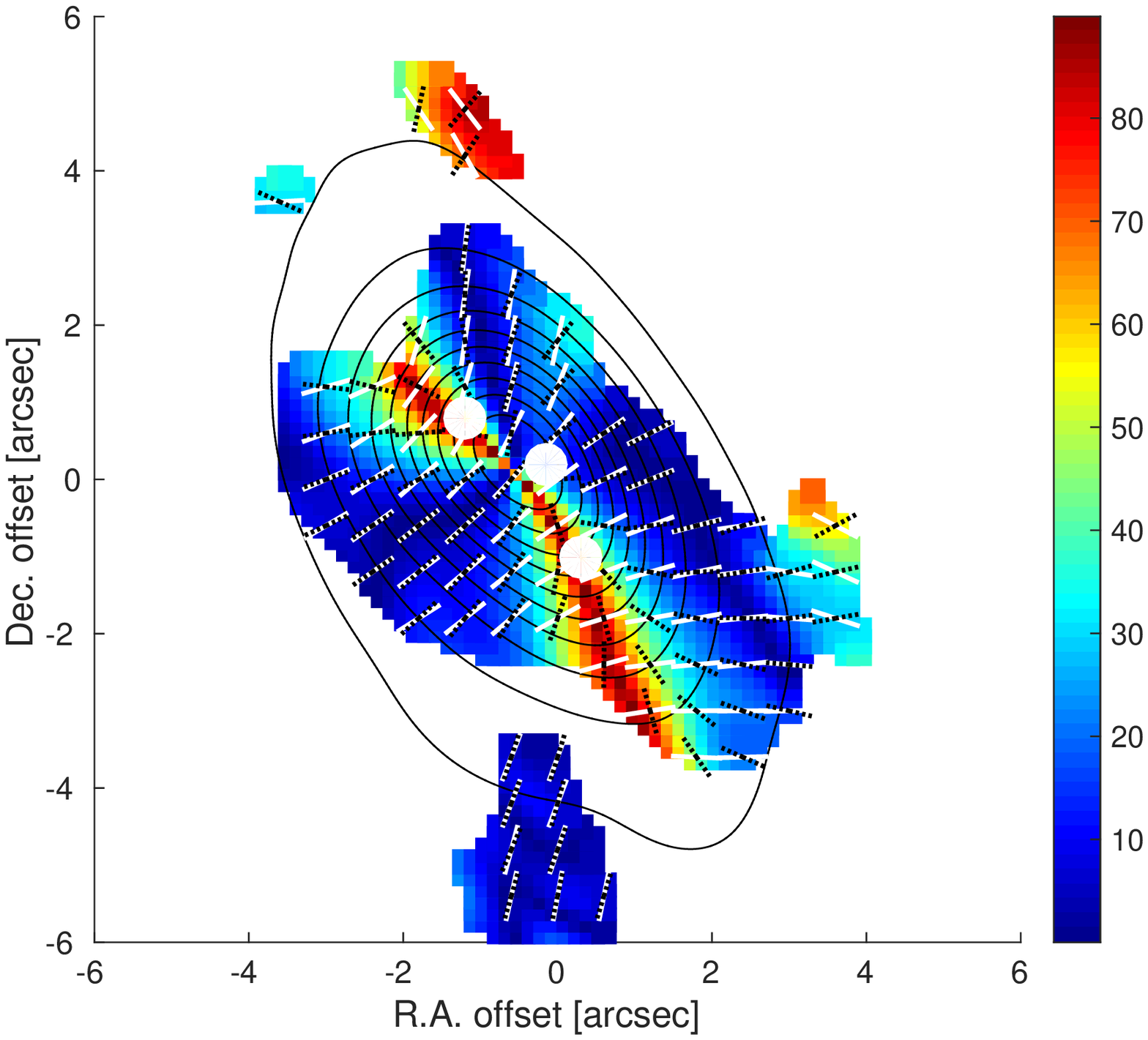}
\includegraphics[scale=0.25]{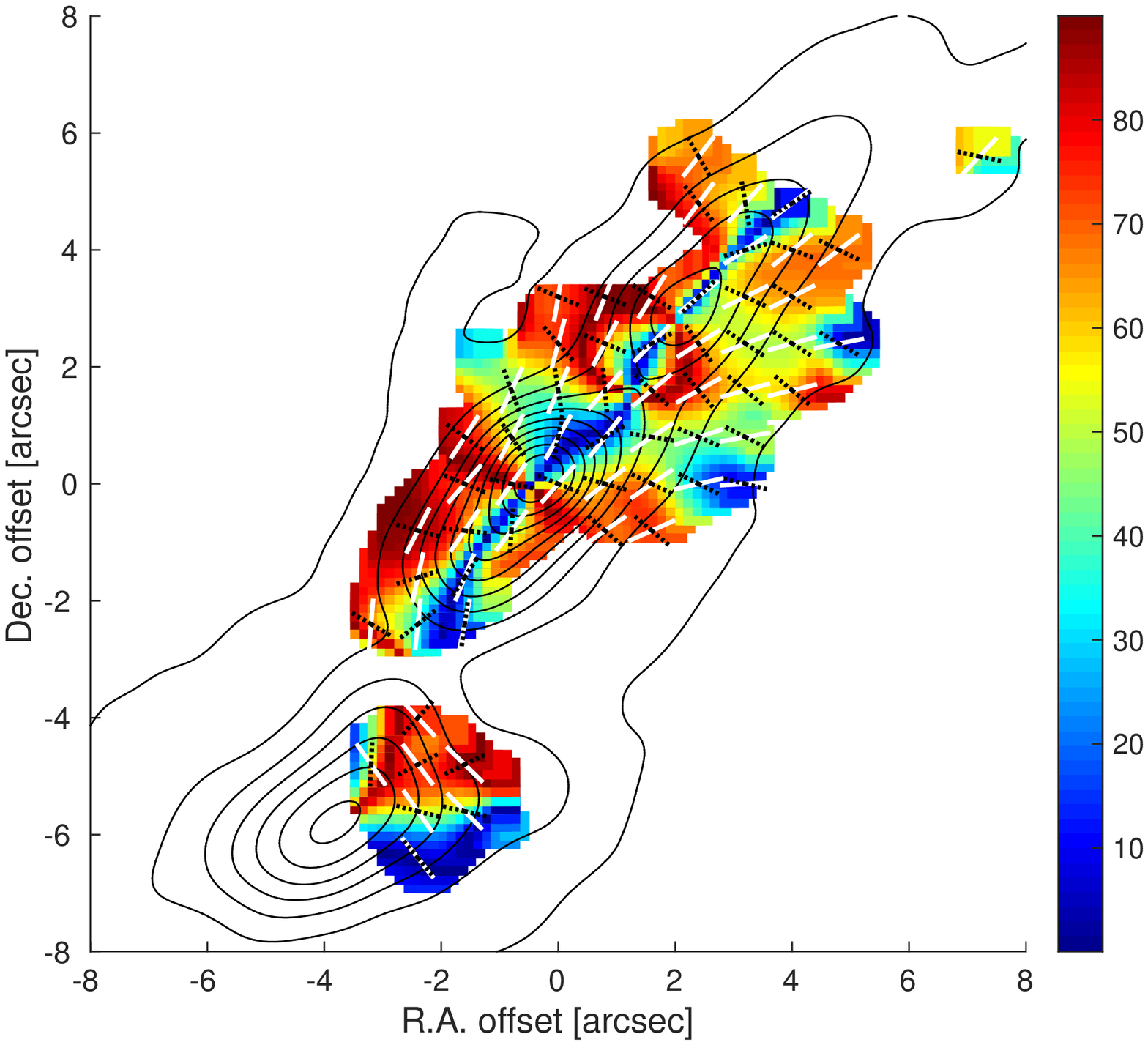}
\caption{\label{example_delta}
Maps of $|\delta|$ (color scale, $0\le |\delta| \le 90^{\circ}$) for G240 (left; SMA, $\theta\sim2''$) and G35.2 (right; SMA, $\theta\sim2''$).
Contours show dust continuum. Magnetic field and emission gradient orientations are displayed with white
and black dotted segments.
The dense condensations (white filled circles) in the center of G240 fall precisely onto regions with large $|\delta|$, towards which material is channeled from the lower $|\delta|$ zones.
}
\end{center} 
\end{figure}

We demonstrate clear systematic features found in $|\delta|$ across two regions observed with the SMA at 345 GHz (Qiu {\em et al.} \cite{qiu13},  \cite{qiu14}) in Figure \ref{example_delta}.
Strikingly, higher-resolution observations
with the SMA at 1.3 mm (Qiu {\em et al.} \cite{qiu09}) show condensations along a mid-plane in G240.31+0.07 where $|\delta|$ is maximised, 
with regions above and below this mid-plane showing small $|\delta|$-values that allow the material to 
be channelled to the mid-plane.  Opposite trends in $|\delta|$ structures are seen in G35.2-0.74N. 
Visually, these sources can clearly be categorised as type IIA (G240) and IIB (G35.2N) as described below.

\begin{figure}
\begin{center}
\includegraphics[scale=0.40]{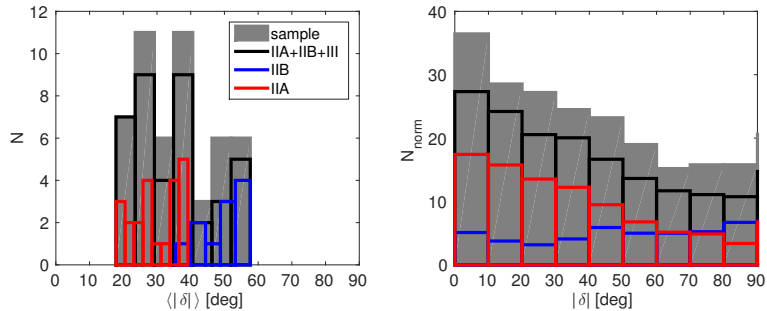}
\caption{\label{hist_delta}
Histograms for the source-averaged $\langle|\delta|\rangle$ (left panel, 50 sources) and $|\delta|$ (right panel, about 
4000 independent measurements from all 50 sources).  Color-coded in red, blue and black are the sources that can
visually easily be identified as type IIA or IIB, and all IIA, IIB and III sources together 
}
\end{center} 
\end{figure}

The examples depicted above are not unique. 
From our sample of 50 targets (SMA, Zhang {\em et al.} \cite{zhang14}; CSO, Dotson {\em et al.} \cite{dotson10})
most of the sources show systematic features in  $|\delta|$ as soon as polarization is detected over a large enough area. 
Our sample covers both low- and high-mass star-formation sites 
over scales of 0.1 to 0.01~pc ($\theta\sim 1-3''$,  $\lambda\sim 870\mu$m; SMA) and larger 
scales around 1pc ($\theta\sim 20''$, $\lambda\sim 350\mu$m; CSO).
Figure \ref{hist_delta} shows histograms for the source-averaged $\langle|\delta|\rangle$ (50 data points)
and for $|\delta|$ (based on 4000 independent beams across all 50 sources). The $|\delta|$-histogram 
of each source is normalized first before co-adding, in order to avoid the statistics being dominated
by sources with the largest numbers of independent beams.  Both histograms show a broad shoulder peaking around 
small $\langle|\delta|\rangle$ and $|\delta|$, falling off with a tail towards larger values. 
This indicates that the prevailing source-magnetic field orientation statistically prefers small misalignments 
with the emission gradient. Thus, magnetic fields are found to be more often within small deviations of a source
minor axis.

\begin{figure}
\begin{center}
\includegraphics[scale=0.36]{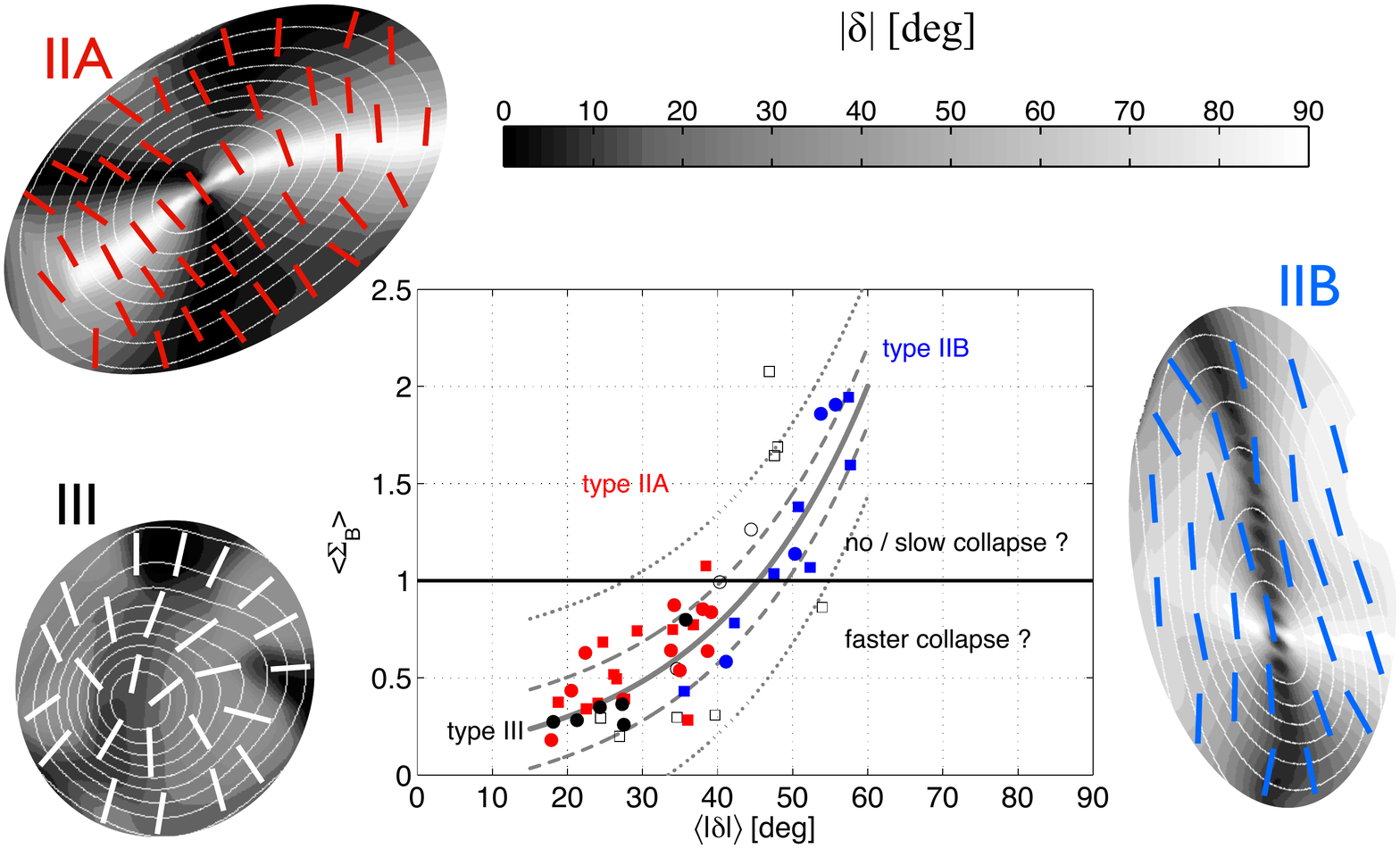}
\caption{\label{sigma_B_vs_delta}
Source-averaged force ratios $\langle \Sigma_B\rangle$ vs source-averaged $\langle |\delta|\rangle$
for the combined SMA and CSO samples. Red and blue filled symbols show
the visually identified IIA- and IIB-type sources from both the SMA (circles) 
and CSO (squares).
Empty symbols correspond to sources with less clear features.
Black filled circles are the visually identified SMA-III sources. 
Schematic magnetic field structures are illustrated for IIA- (red segments), 
IIB- (blue segments) and III-type sources (white segments).  White contours display
the dust continuum emission.  The black-to-white color grading on the top indicates the local $|\delta|$.
Clear opposite trends are apparent for type-IIA and -IIB sources.
The solid gray line is the log-linear best fit 
$\langle\Sigma_B\rangle = A\cdot \exp(B\cdot \langle|\delta|\rangle)$ with 
$A=0.116$ and $B=0.047$. The dashed and dotted gray lines indicate mean error
($\pm \,0.20$) and $3\sigma$ ($\pm \, 3\cdot 0.19$) bounds.
}
\end{center} 
\end{figure}

If $|\delta|$ is further interpreted with an MHD force equation, the ratio 
between the magnetic field tension force and gravity, $\Sigma_B$, can be calculated from $|\delta|$
and $\psi$ as $\Sigma_B=\sin\psi / \sin(\pi/2-|\delta|)$, where $\psi$ is the local angle between the 
gradient of the dust continuum and the direction of local gravity (Koch {\em et al.\/} \cite{koch12a}). 
This finding reflects the 
idea that an observed molecular cloud morphology -- with its characteristic angles $\delta$
and $\psi$ -- is a measure of the imprint of all the relevant forces in a system. 
Figure \ref{sigma_B_vs_delta} displays source-averaged force ratios $\langle\Sigma_B\rangle$
versus source-averaged $\langle|\delta|\rangle$. The force ratio increases with 
growing $\langle|\delta|\rangle$. 

\section{Conclusion}

The local angle $|\delta|$ is found to be a prime observable across a sample of 50 star-forming 
regions.  It can further be linked to a magnetic field-to-gravity force ratio, $\Sigma_B$.
Maps of $\Sigma_B$ are closely correlated to features in $|\delta|$-maps, making $|\delta|$
a tracer for the local role of the magnetic field.  While Figure \ref{sigma_B_vs_delta}
shows a source-averaged correlation, it is important to note that our approach also provides a local 
assessment of the magnetic field. Thus, $|\delta|$ can explain where condensations, and 
potentially star formation, occur (Figure \ref{example_delta}). Our finding of magnetic fields
being more aligned parallel to emission gradients in clouds and cores (Figure \ref{hist_delta}) connects to 
{\it Planck} results that observe a transition from fields being parallel to ridges in the diffuse
interstellar medium to more orthogonal as densities towards molecular clouds increase 
(Planck Collaboration XXXII \cite{planckxxxii}).


\end{document}